\documentclass[a4paper,aps,pre,superscriptaddress,floatfix,nofootinbib,twocolumn,showpacs]{revtex4-1}

\usepackage[pdftex]{graphicx}
\usepackage{latexsym,amsmath}
\usepackage[pdftex]{color}
\usepackage{hyperref}
\usepackage{url}

\newcommand{\av}[1]{\langle {#1} \rangle}

\newcommand{\FigPath}{.}

\begin{document}

\title{Model reproduces individual, group and collective dynamics \\
of human contact networks}

\author{Michele Starnini} 
  \address{Departament de F\'\i sica i Enginyeria Nuclear,
  Universitat Polit\`ecnica de Catalunya, Campus Nord B4, 08034
  Barcelona, Spain}
  
\author{Andrea Baronchelli} 
\address{Department of Mathematics, City University London, Northampton
  Square, London EC1V 0HB, UK}

\author{Romualdo Pastor-Satorras} 
\address{Departament de F\'\i sica i Enginyeria Nuclear,
  Universitat Polit\`ecnica de Catalunya, Campus Nord B4, 08034
  Barcelona, Spain}
  

\begin{abstract}
  Empirical data on the dynamics of human face-to-face interactions 
  across a variety of social venues 
  have recently revealed a number of
  context-independent structural and temporal properties of human
  contact networks. This universality
  suggests that some basic mechanisms may be responsible for the unfolding 
  of human interactions in the physical space. Here we
  discuss a simple model that reproduces the empirical distributions for
  the individual, group and collective dynamics of face-to-face contact
  networks. The model describes agents that move randomly in a
  two-dimensional space and tend to stop when meeting `attractive'
  peers, and reproduces accurately the empirical distributions. 
\end{abstract}

\maketitle

\section{Introduction}
\label{sec:intro}

Social and Cognitive Sciences have experienced a major transformation in the
past few years \cite{lazer2009life,watts2007twenty,baronchelli2013networks}.  The recent
availability of large amounts of data has indeed fostered the
quantitative understanding of many phenomena that had previously been
considered only from a qualitative point of view
\cite{barabasi2010bursts,jackson2010social}. Examples range from human
mobility patterns \cite{brockmann2006scaling}
and human behavior in economic arenas
\cite{radicchi2012rationality,preis2013quantifying}, to the analysis of
political trends
\cite{adamic2005political,carpenter2004friends,lazer2011networks}. Together with the
World-Wide Web, a wide array of technologies have also contributed to
this data deluge, such as mobile phones or GPS devices
\cite{eagle2009inferring,takhteyev2012geography,mocanu2013twitter},
radio-frequency identification devices
\cite{10.1371/journal.pone.0011596}, or expressly designed online
experiments \cite{centola2010spread}. The understanding of social networks  has clearly benefitted from this trend \cite{jackson2010social}. Different large social networks, such as mobile phone  \cite{onnela2007structure} or email  \cite{bird2006mining} communication networks, have been characterized in detail while the rise of online social networks has provided an ideal playground for researchers in the social sciences \cite{huberman2008social,kwak2010twitter,ellison2007benefits}. The availability of data, finally, has allowed to test the validity of the different models of social networks that have mainly been published within the physics-oriented complex networks literature, bridging the gap between mathematical speculations and the social sciences \cite{toivonen2009comparative}.

Among the different kinds of social networks, a prominent position is occupied by the so-called face-to-face contact networks, which represent a pivotal substrate for the
transmission of ideas \cite{nohria2000face}, the creation of social
bonds \cite{storper2004buzz}, and the spreading of infectious diseases
\cite{liljeros2001web,salathe2010high}. The uniqueness of these networks stems from the fact that face-to-face conversation is considered the ``gold standard'' \cite{nardi2002place} of communication \cite{clark1991grounding,kiesler1984social}, and although it can be costly \cite{hollan1992beyond}, the benefits it contributes to workplace efficiency or in sustaining social relationships are so-far unsurpassed by the economic convenience of other forms of communication \cite{nardi2002place}. It is because face-to-face interactions bring about the richest information transfer \cite{doherty1997face}, for example,  that in our era of new technological advancements business travel has kept increasing so steadily \cite{nardi2002place}. In light of all this, it is not surprising that face-to-face interaction networks have long been the focus of a major attention \cite{bales1950interaction,bion2013experiences,arrow2000small}, but the lack of fine-grained and time-resolved data represented a serious obstacle to the quantitative comprehension of the dynamics of human contacts. Researchers in social network analysis had in fact long pointed out the importance of the temporal dimension for the understanding of social networks, which are not static entities, but rather vary in time \cite{carley1991theory,doreian1996brief,moody2002importance}.


Recently, the so-called data revolution has invested also the study of human face-to-face contact networks  \cite{sociopatterns,stopczynski2014measuring}. 
In particular, the fine-grained measurement of face-to-face interactions using wearable active radio-frequency identification devices (RFID), performed by the SocioPatterns collaboration \cite{sociopatterns},  revealed the complex temporal
and structural properties of human contact networks
\cite{10.1371/journal.pone.0011596}. 
Important among these properties is
the bursty nature of human social contacts \cite{barabasi05}, revealed
in the distributions of the time of contact between pairs of
individuals, the total time spent in contacts by a given individual, or
the inter-event times between consecutive contacts involving the same
individual, all exhibiting heavy tails, more or less compatible with a
power-law form
\cite{10.1371/journal.pone.0011596,PhysRevE.85.056115}. The fact that
these statistical regularities are common to such
apparently diverse settings as schools, hospitals, scientific
conferences, and museums, suggests that the properties of human
face-to-face contact networks can be explained by some fundamental,
general process, independently from the considered situation, and 
calls for simple models to explain and reproduce these features
\cite{Isella:2011qo,zhao2011social}.

In this paper we present and analyze a simple model able to replicate
most of the main statistical regularities exhibited by human
face-to-face contact networks data. The key insight of the model is the
suggestion that the social ``attraction'' of individuals may be the
major responsible for the observed phenomenology of face-to-face contact
networks. This insight is implemented by allowing individuals, each
characterized by an intrinsic social \textit{attractiveness}, to wander
randomly in a two dimensional space---representing the simplified
location of a social gathering---until they meet someone, at which point
they have the possibility of stopping and starting a ``face-to-face''
interaction. Without entering into the problem of the definition of
attractiveness, we adopt here an operative approach: Attractive
individuals are more likely to make people stop around them and start an interaction, but they
are also more prone to abandon their interactions if these are initiated
by less attractive agents.

We observe that these simple rules, and the asymmetry of the
interactions that they imply, are sufficient to reproduce quantitatively
the most important features of the empirical data on contact
networks. We explore in particular properties belonging to three
different scales. At \textit{the individual, or `microscopic', level}, we focus in
temporal properties related to the distributions of contact durations or
inter-contact times, and in structural properties related to the time
integrated representation of the contact data, as usually reported in
the literature \cite{jackson2010social}. A preliminary
account of the model results at this level was presented in
Ref.~\cite{PhysRevLett.110.168701}. Moving beyond the analysis of
individual properties, we consider here \textit{the group, or `mesoscopic', level}, represented by
groups of simultaneously interacting individuals, which typify a crucial
signature of face-to-face networks \cite{bales1950interaction,arrow2000small} and have important consequences on
processes such as decision making and problem solving
\cite{buchanan07}. We measure the distribution of group
sizes as well as the distribution of duration of groups of different
size. We finally zoom one more step out and inspect the 
\textit{collective, or `macroscopic', level} looking at properties
that depend on the time interaction pattern of the whole population. We
address in particular the issue of the causality patterns of the
temporal network, as determined by the time-respecting paths between
individuals \cite{moody2002importance,PhysRevE.71.046119} and the network reachability,
defined as the minimum time for information to flow from an individual
$i$ to another individual $j$ and measured by means of a searching
process performed by a random walker
\cite{moody2002importance,PhysRevE.85.056115,perra_random_2012}. We observe that the model
 reproduces not only qualitatively, but also quantitatively, the
properties measured from empirical data at all the 
scales. 

Finally, as a check for robustness, we explore the role of the the
parameters that define the model. Particular emphasis is made on the
motion rule adopted by the individuals. While a simple random walk for
the individuals' movements is initially considered, in fact, a
consistent amount of literature suggests that L\'evy flights
\cite{viswanathan2011physics} might provide a better characterization of
human movement
\cite{brockmann2006scaling,gonzalez2008understanding,rhee2011levy,baronchelli2013levy}.
We observe that the results of the model are robust with respect to
various possible alterations of the original formulation, including the
adopted rule of motion.

The paper is structured as follows: In Section~\ref{sec:data} we discuss
the SocioPatterns experiment, and present the time-varying network
formalism used to represent its data.  The model is defined in detail in
Section~\ref{sec:model}, while Section~\ref{sec:indiv},
Section~\ref{sec:group} and Section~\ref{sec:collective}, address the
model behavior concerning the individual, group and collective dynamics,
respectively.  In Section~\ref{sec:robustness} we show the model
robustness with respect to the variation of the main parameters
involved. Finally, Section~\ref{sec:discussion} is devoted to
discussion, with particular attention to the crucial role of social
attractiveness in the model.

\section{Empirical data}
\label{sec:data}

\subsection{The SocioPatterns experiment}
\label{sec:sociopatterns-data}

Here we consider the data on human contact networks as recorded by the
SocioPatterns collaboration \cite{sociopatterns} in closed gatherings of
individuals, covering scientific conferences, hospital wards, and
schools. In the deployments of the SocioPatterns infrastructure, each
individual participating in the experiment wears a badge equipped with
an active radio-frequency identification (RFID) device, able to relay
the information about the close proximity of other devices. The
emissions of the RFIDs are of low frequency and power, and highly
directional. Thus, the close
proximity registered by two devices can be associated to their respective wearers being face
to face at a short distance, a fact that can be assumed to correspond to
a conversation taking place among them.  The devices properties are
tuned so that face-to-face interactions are recorded with a space
resolution of $1-2$ meters and a time resolution $\Delta t_0 \sim 20$
seconds, representing  the elementary time interval in the contact
network evolution.

We will contrast numerical simulations of the proposed model against
empirical results from SocioPatterns deployments in several different
social contexts: a Lyon hospital (``Hospital"), the Hypertext 2009
conference (``Conference"), the Soci\'et\'e Francaise d'Hygi\'ene
Hospitali\'ere congress (``Congress") and a high school (`School"). A
further description of these datasets can be found in
\cite{10.1371/journal.pone.0011596,Isella:2011qo,Stehle:2011,percol}.

\subsection{Face-to-face networks as time-varying graphs}
\label{sec:tempnet}

The empirical data collected by the SocioPatterns deployments are
naturally described in terms of temporally evolving graphs (temporal
networks) \cite{moody2002importance,Holme2012}, whose nodes are defined by a static
collection of individuals, and whose links represent pairwise
interactions, which appear and disappear over time. Interactions between
individuals are aggregated over a time window $\Delta t_0 = 20$s,
corresponding to the natural resolution of the RFID devices. Thus, all
the interactions established within this time interval are considered as
simultaneous and contribute to build a ``instantaneous'' contact
network, formed by isolated nodes and small groups of interacting
individuals. Therefore, each dataset consists of a time-varying network
with a number $N$ of different interacting individuals and a total
duration of $T$ elementary time steps. This procedure is standard in the
study of time-varying networks, and represents a tractable and good
approximation as far as the aggregation window is not too large
\cite{ribeiro2013quantifying}.

An exact representation of temporal networks is given in terms of a
\emph{contact sequence}, representing the sequence of contacts (edges)
as a function of time. This sequence is expressed by a characteristic
function \cite{Newman2010}, $X(i,j,t)$, taking the value $1$ when nodes
$i$ and $j$ are connected at time $t$, and zero otherwise, for $t \in
[0,T]$.  The temporal patterns of the contact sequence can be
statistically characterized as its most basic (individual) level by the
distribution of the duration $\Delta t$ of contacts (conversations)
between pairs of individuals, $P(\Delta t)$, and the distribution of gap
times, $\tau$, between two consecutive conversations involving a common
individual and two other different individuals, $P(\tau)$.

A coarse-grained view of temporal networks can by obtained by means of a
projection into an aggregated static network, integrated over the whole
observation time $T$.
The edges in the integrated networks indicate the presence of a contact
between two nodes at any point in the past. One key variable that characterizes the topology of the network is the degree of a node $k_i$,
is  associated to the total number of different contacts node $i$
has had during the integration time $T$, which can be interpreted as a measure of social integration \cite{marsden1987core} or activity \cite{wasserman1994social}.
Edges can be additionally classified according to their importance or role (eg., family, friends, work colleagues, acquaintances) \cite{granovetter1973strength,granovetter1973strength,wellman1990different}, and heterogeneities in links can be revealed also in the specific case of face-to-face networks. In particular, each edge is annotated with a weight $w_{ij}$, indicating the total time spent in
interactions by the pair of nodes $i$ and $j$, and which is defined by
\begin{equation}
  \label{eq:2}
  w_{ij} = \dfrac{1}{T}\sum_{t} X(i,j,t).
\end{equation}
The strength of node $i$, defined as $s_i = \sum_j w_{ij}$, expresses
then the cumulative time spent in interactions by individual $i$
\cite{onnela2007structure,Isella:2011qo,Stehle:2011}.  The aggregated
representation is an useful benchmark to point out the effect of
temporal correlations \cite{PhysRevE.85.056115}, and it allows to
identify interesting properties of the system. For example, the observed super-linear relation between the degree and strength of an individual implies
that on average highly-connected individuals spend more time in each
interaction with respect to the poorly-connected ones
\cite{10.1371/journal.pone.0011596,baronchelli2013networks}. More in general, the strength of links and individuals helps not only to understand the structure of a social network, but also the dynamics of a wide range of phenomena involving human behavior, such as the formation of communities and the spreading of information and social influence \cite{hill2010infectious,onnela2007structure,watts2007twenty}

A summary of the average properties of the datasets considered in this
work is provided in Table~\ref{tab:summary}. The statistical and
structural properties of the temporal networks representing
SocioPatterns data have been extensively studied in the literature, see
Refs.~\cite{10.1371/journal.pone.0011596,Isella:2011qo,PhysRevE.85.056115}.
We will review them in the following sections, when performing a
numerical comparison with the outcome of the proposed model.

\begin{table}[t]
    \begin{tabular}{|c||c|c||c|c||c|c|}
    \hline
    Dataset & $N$ & $T$  & $\overline{p}$ &$\av{\Delta t}$ & $\av{k}$ & $\av{s}$ \\ \hline  
      Hospital       &    84   & 20338 & 0.049  & 2.67 & 30 & 0.0563 \\ 
      Conference   &    113 & 5093   & 0.060  & 2.13 & 39 & 0.0719 \\ 
      School          &  126   & 5609   & 0.069  & 2.61 & 27 & 0.0808 \\  
      Congress     &  416   & 3834   & 0.075  & 2.96 & 54 & 0.131 \\ \hline 
    \end{tabular}
    \caption{Some properties of the SocioPatterns datasets under
      consideration: $N$, number of different individuals engaged in
      interactions; $T$, total duration of the contact sequence, in units of
      the elementary time interval $t_0 = 20$ seconds; $\overline{p}$, average
      number of individuals interacting at each time step; $\av{\Delta t}$,
      average duration of a contact; $\av{k}$ and $\av{s}$: average degree and
      average strength of the projected network, aggregated over the whole
      sequence (see main text).} \label{tab:summary} 
\end{table}



\section{A model of social attractiveness}
\label{sec:model}

The model we propose in defined as follows
\cite{PhysRevLett.110.168701}: $N$ individuals, free to move in a closed
environment, can interact between them when situated within a small
distance (that we assimilate to the exchange range of the RFID
devices). Agents perform a random walk of fixed step length $v$ in a
closed box of linear size $L$ (corresponding to a density $\rho= N/L^2$)
and start interacting whenever they intercept another agent within a
certain distance $d$. Crucially, each agent is characterized by an
intrinsic ``attractiveness'' or social appeal, a consequence of their
social status or the role they play in social gatherings, and which
represents her power to raise interest in the others. The attractiveness
$a_i$ of each individual is randomly chosen from a given distribution
$\eta(a)$, bounded in the interval $a_i \in [0,1]$, which we choose to
be uniform. Thus, when interacting, an individual either interrupts his
motion to preserve an interaction, and this happens with a probability
proportional to the attractiveness of the most interesting neighbor, or
keeps moving. This translates into a walking probability $p_i(t)$ of the
agent $i$ of the form
\begin{equation}
\label{eq:mr}
p_i(t) = 1- \max_{j \in \mathcal{N}_i(t) } \{ a_j \} ,
\end{equation}
where $\mathcal{N}_i(t)$ is the set of neighbors of agent $i$ at time
$t$. Therefore, the more attractive an agent $j$ is, the more interest
she will raise in a neighbor agent $i$, who will slow down her random
walk exploration accordingly.

Empirical data show that not all the agents are simultaneously present
in the system, but they can be in an active state, moving and
interacting, or in an inactive state, without moving nor interacting. To
take this fact into consideration, each agent $i$ in the model is
further characterized by an activation probability $r_i$. At each time
step, one inactive agent $i$ can become active with a probability $r_i$,
while one active and isolated agent $j$ can become inactive with
probability $1 - r_j$. We will choose the activation probability $r_i$
of the agents randomly from an uniform distribution $\phi(r)$, bounded
in $r_i \in [0,1]$.

The results presented have been numerically simulated adopting the
parameters $v=d=1$, $L=100$ and $N=200$, for a total duration $T= 2
\times 10^4$ time steps, unless otherwise specified.  In the initial
conditions, agents are placed at randomly chosen positions, and are
active with probability $1/2$.  Numerical results are averaged over
$10^2$ independent runs.

Before proceeding a comment is in order. We adopt here an operational
definition of ``attractiveness" as the property of an individual to
attract the interest of other individuals, making them willing to engage
in a conversation, or to listen to what he/she is saying. Thus, we do
not enter in any speculations on what are the cultural or psychological
factors that make a person attractive in this sense, but we reckon that
many possible candidates exist, ranging from status
\cite{hollingshead1975four} to extroversion
\cite{scherer1978personality}. In light of the success of our model in
reproducing the empirical distributions (see
below), we consider that identifying which feature, or set of features,
the attractiveness is a proxy of represents one important direction for
future work.

\section{Individual level dynamics}
\label{sec:indiv}

In this section we compare the individual level predictions of the model
against the observations from empirical data.

\subsection{Temporal correlation}

The temporal pattern of the individual contacts is one of the most
distinctive feature of face-to-face interaction networks
\cite{10.1371/journal.pone.0011596,PhysRevE.85.056115}, and in general
of time-varying networks \cite{moody2002importance}.  Relevant quantities
measuring the correlation of this temporal pattern are \cite{Holme2012}:

\begin{itemize}
\item the distribution of the duration $\Delta t$ of the contacts
  between pairs of agents, $P(\Delta t)$; 
\item the distribution of gap times $\tau$ between two consecutive
  conversations involving a common individual, $P(\tau)$.
\end{itemize}

\begin{figure}[tb]
   \centerline{\includegraphics*[width=0.90\linewidth]{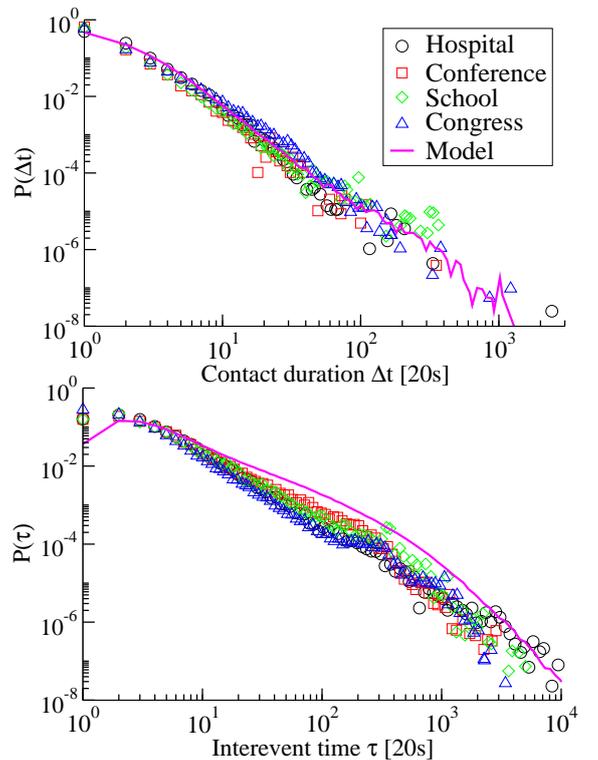}}
   \caption{(color online) {\bf Burstiness of human contacts:}
     Distribution of the contact duration, $P(\Delta t)$ (top) and the
     time interval between consecutive contacts, $P(\tau)$, (bottom) for
     various datasets (symbols) and for the attractiveness model (line).}
     \label{fig:t_distr} 
 \end{figure}

Empirical data reveal that both distributions are broad, with long tails
that can be described in terms of a power-law function
\cite{10.1371/journal.pone.0011596}, reflecting the bursty nature of human
interactions \cite{Oliveira:2005fk,barabasi05}. Most contacts are short
and separated by small amounts of time, but long interactions, as well
as long inter-contact times, are always possible.
Figure~\ref{fig:t_distr} shows that the model captures this property and
quantitatively reproduces the phenomenology of the various datasets. 


 \begin{figure}[tb]
    \centerline{\includegraphics*[width=0.90\linewidth]{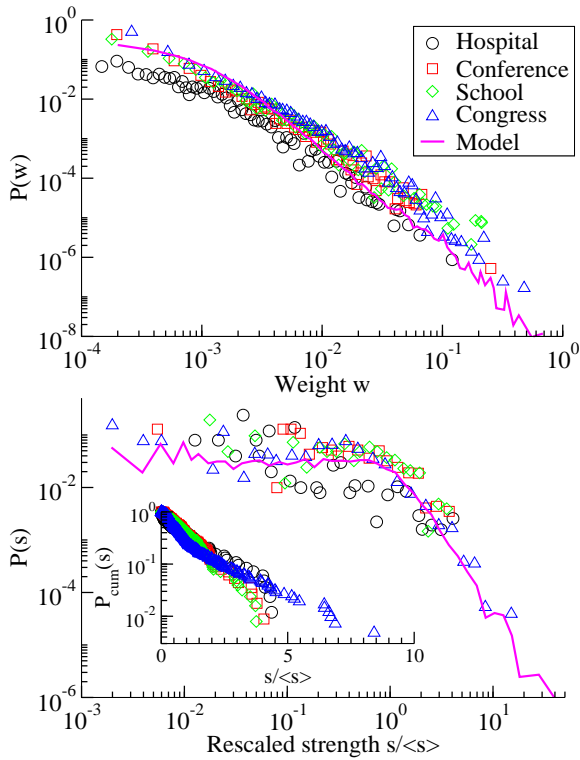}}
   \caption{(color online) {\bf Time-integrated network: } 
    Distribution of the weight, $P(w)$, (top) and rescaled strength, $P(s)$ (bottom), 
     for various datasets (symbols) and for the attractiveness model (line). 
     In the inset we plot the cumulative strength distribution, $P_{\rm{cum}}(s)$, 
     to highlight the exponential decay of the $P(s)$ (see Section \ref{sec:discussion}).}
     \label{fig:w_distr} 
 \end{figure}
 
\subsection{Time-aggregated networks}

Additional information regarding the pattern of individual interactions
is obtained by integrating the time-varying network into an aggregated
weighted network. As mentioned in section \ref{sec:tempnet}, this
corresponds to considering all the interactions occurring in a given
time window $\Delta t$ in the limit of $\Delta t \rightarrow T$,
i.e. all the interactions taking place in the dataset.  Repeated
contacts between any two individuals $i$ and $j$ increase the weight
$w_{ij}$ associated to the link connecting them, and the sum of the
weights of the links connecting an individual $i$ to his peers (i.e. the
total time $i$ spends in conversations) represents the strength $s_i$.

In Fig. \ref{fig:w_distr} we plot the distribution $P(w)$ of weights
$w_{ij}$ between pairs of agents, and the cumulative distribution of the
strength, $P(s)$, showing that the numerical simulation of the model are
in excellent agreement with empirical data.  The heavy tailed weight
distribution, $P(w)$, shows that the heterogeneity in the duration of
individual contacts persists even when interactions are accumulated over
longer time intervals.  The strength distribution $P(s)$ has instead an
exponential tail, as revealed by the cumulative distribution
$P_{\rm{cum}}(s)$, defined as the probability of finding any individual
with a strength larger than $s$, see inset of Fig.~\ref{fig:w_distr}. 
In both cases, our model leads to results that are fully compatible with
the empirical evidence.

\section{Group level dynamics}
\label{sec:group}

\begin{figure}[t]
  \centerline{\includegraphics[width=0.90\linewidth]{\FigPath/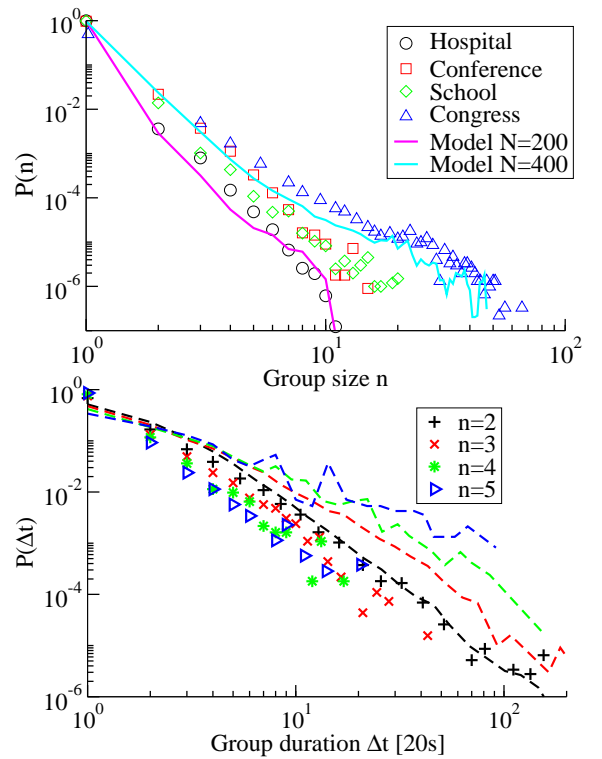}}
  \caption{\label{fig:gr_distr} (color online) {\bf Group dynamics:} Up:
    Group size distribution $P(n)$ for different datasets and for the
    model, numerically simulated with different number of agents $N=200$
    and $N=400$, and same size $L=100$.  Down: Lifetime distribution
    $P_n(\Delta t)$ of groups of different size $n$, for the ``Congress"
    dataset (symbols) and for the model numerically simulated with
    $N=400$ and $L=100$. (dashed lines).  }
  \end{figure}

Another important aspect of human contact networks is the dynamics of
group formation \cite{bales1950interaction,arrow2000small}, defined by a set of $n$ individuals interacting
simultaneously, not necessarily all with all, for a period of time
$\Delta t$. As we have noted above, such groupsplay the important role of catalysts for decision making and problem
solving \cite{buchanan07}.  In Fig. \ref{fig:gr_distr} (top) we plot the
probability distribution of observing a group of size $n$, $P(n)$, in
any instant of the ongoing social event, for the different empirical
data sets.  The distributions are compatible with a power law behavior,
whose exponent depends on the number of agents involved in the social
event, with larger datasets (such for example the Congress one, see
Table \ref{tab:summary}) capable of forming bigger groups with respect
to smaller data sets.  Clearly, the model predictions are in substantial
agreement with the data when we inform the model with a sensible,
data-driven, value of $N$.

In order to explore the dynamics of group formation, we define the
lifetime $\Delta t$ of a group of size $n$ as the time spent in
interaction by the same set of $n$ individuals, without any new arrival
or departure of members of the group.  In Fig. \ref{fig:gr_distr} (bottom)
we plot the lifetime distribution $P_n(\Delta t)$ of groups of different
sizes $n$, finding that experimental and numerical results have a
similar power-law behavior.  We note however that for empirical data the
lifetime distribution $P_n(\Delta t)$ decays faster for larger groups,
i.e. big groups are less stable than small ones, while the model outcome
follows the opposite behavior.  This means that, in the model, larger
groups are (slightly) more stable than observed in the data. This is
probably due to the fact that, the larger the group, the bigger the
probability of finding two (or more) individuals with large
attractiveness in the group, which guarantee the stability against
departures. However, an alternative explanation could be that the
RFID devices of the SocioPatterns experiment require individuals to
face each others within a given angle, making the group
definition effectively more fragile than in the model, where such directionality is absent. 

\section{Collective level dynamics and searching efficiency}
\label{sec:collective}


The temporal dimension of any time-varying graph has a deep influence on
the dynamical processes taking place upon such structures \cite{moody2002importance}. In the
fundamental example of opinion (or epidemic) spreading, for example, the
time at which the links connecting an informed (or infected) individual
to his neighbors appear determines whether the information (or
infection) will or will not be transmitted. In the same way, it is
possible that a process initiated by individual $i$ will reach
individual $j$ through an intermediate agent $k$ through the path $i
\rightarrow k \rightarrow j$ even though a direct connection between $i$
and $j$ is established later on. This information is lost in the time
aggregated representation of the network, where any two neighboring
nodes are equivalent \cite{Holme2012}. In general, time respecting paths
\cite{PhysRevE.71.046119} determine the set of possible causal
interactions between the agents of the graph, and the state of any node
$i$ depends on the state of any other vertex $j$ through the collective
dynamics determining their causal relationship.

\begin{figure}[tb]
  \centerline{\includegraphics[width=0.90\linewidth]{\FigPath/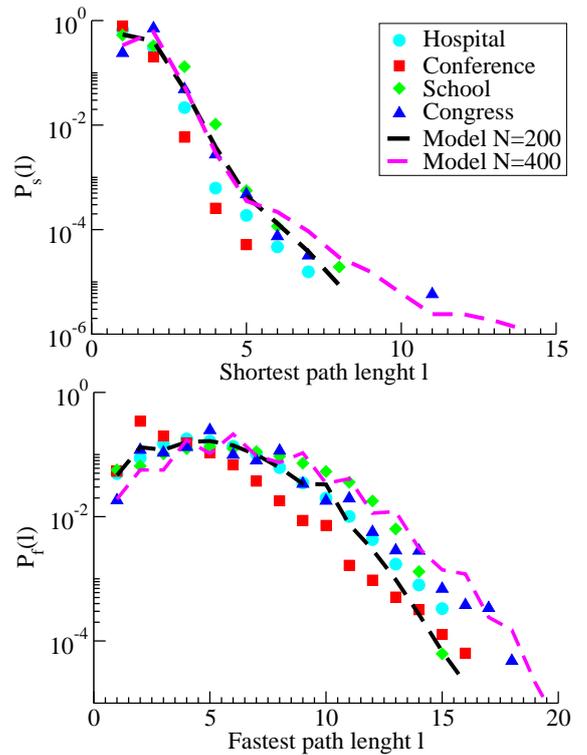}}
  \caption{\label{fig:short} (color online) {\bf Time-respecting paths:}
    Probability distributions of the shortest, $P_s(l)$, (top) and
    fastest, $P_f(l)$, (bottom) path length, for the time-varying network
    obtained by the empirical data and by the model, numerically
    simulated with different number of agents $N=200$ and $N=400$, and
    same size $L=100$. }
\end{figure}

For any two vertices, we can measure the shortest time-respecting path,
$l_{ij}^{s}$, and fastest time-respecting path, $l_{ij}^{f}$, between
them \cite{PhysRevE.85.056115}. The former is defined as the path with
the smallest number of intermediate steps between nodes $i$ and $j$, and
the latter is the path which allows to reach $j$ starting from $i$
within the smallest amount of time.  In Fig. \ref{fig:short} we plot the
probability distributions of the shortest and fastest time-respecting
path length, $P_s(l)$ and $P_f(l)$, respectively, of both empirical data
and model, finding that they show a similar behavior, decaying
exponentially, and being peaked for a small number of steps.

Given the importance of causal relationship on any spreading dynamics,
it is interesting to explicitly address the dynamical unfolding of a
diffusive process. Here we analyze the simplest example of a search
process, the random walk, which describes a walker traveling the network
and, at each time step, selecting randomly its destination among the
available neighbors of the node it occupies. The random walk represents
a fundamental reference point for the behavior of any other diffusive
dynamics on a network, when only local information is available. Indeed,
assuming that each individual knows only about the information stored in
each of its nearest neighbors, the most naive economical strategy is the
random walk search, in which the source vertex sends one message to a
randomly selected nearest neighbor \cite{PhysRevE.64.046135}.  If that
individual has the information requested, it retrieves it; otherwise, it
sends a message to one of its nearest neighbors, until the message
arrives to its final target destination.  In this context, a quantity of
interest is the probability that the random walk actually find its
target individual $i$ at any time in the contact sequence, $P_r(i)$, or
\emph{global reachability} \cite{moody2002importance,PhysRevE.85.056115}.

\begin{figure}[tb]
  \centerline{\includegraphics[width=0.90\linewidth]{\FigPath/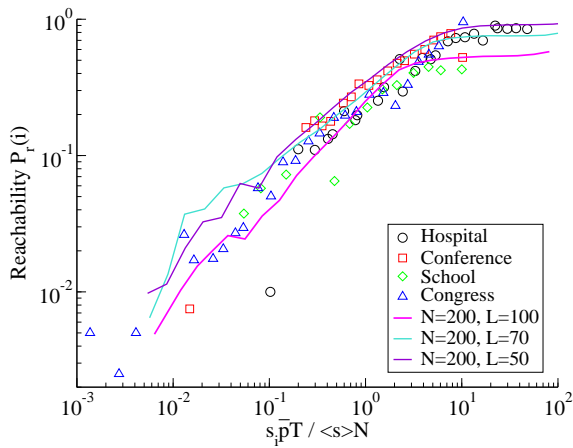}}
  \caption{\label{fig:prob_reach} (color online) {\bf Reachability:}
    Correlation between the reachability of agent $i$, $P_r(i)$, and his
    rescaled strength, $ s_i \bar{p}T / \av{s}N$.  The empirical data
    sets considered (symbols) and the model (lines), numerically
    simulated with different density $\rho$, follow a close behavior.
    We averaged the reachability of each individual over at least $10^2$
    different runs, starting with different source node.  }
\end{figure}

In principle, the reachability of an individual $i$ must be correlated
with the total time spent in interactions, namely his strength $s_i$,
but it also depends on the features of the considered social event, such
as the density of the interaction $\bar{p}$, the total duration $T$ and
possibly other event-specific characteristic (see Table
\ref{tab:summary} for information of the different data sets
considered).  On the basis of a simple mean field argument, it has been
shown \cite{PhysRevE.85.056115} that the probability of node $i$ to be
reached by the random walk, $P_r(i)$, is correlated with its relative
strength $s_i/\av{s}N$, times the average number of interacting
individuals at each time step, $\bar{p}T$.  In Fig \ref{fig:prob_reach}
we plot the reachability $P_r(i)$ against the rescaled strength $ s_i
\bar{p}T / \av{s}N$, showing that very different empirical data sets
collapse into a similar functional form.  Remarkably, the model is able
to capture such behavior, with a variability, also found in the data,
which depends on the density $\rho$ of the agents involved.  As noted
for the group dynamics, a larger density corresponds to a higher
reachability of the individuals.  We note that the empirical data are
reproduced by the model for the same range of density considered in the
previous analysis.

\section{Model robustness}
\label{sec:robustness}

\begin{figure}[tb]
  \centerline{\includegraphics[width=0.90\linewidth]{\FigPath/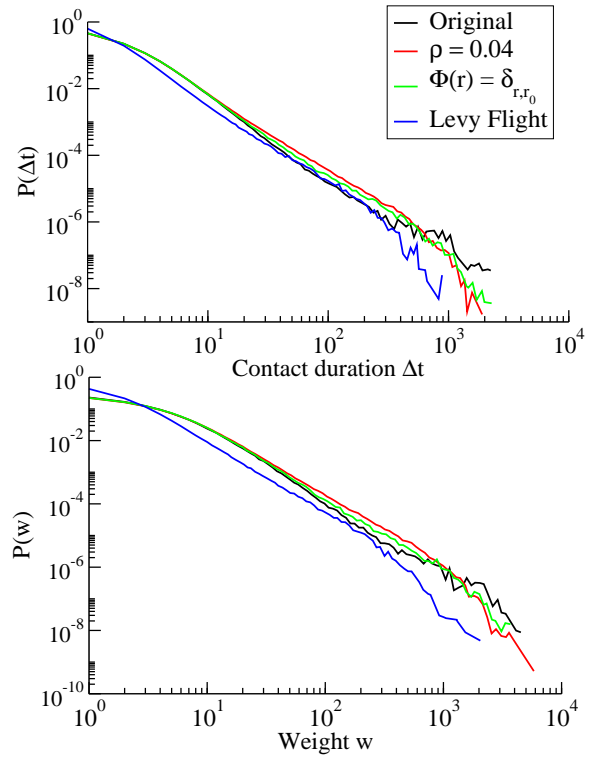}}
  \caption{\label{fig:robust} (color online) {\bf Robustness:}
  Contact duration (top) and weigh (bottom) probability distributions
  obtained by simulating the model in its original definition, 
  with a different density $\rho$ given by $N=400$ individuals,
  with constant activation probability $\Phi(r)=\delta_{r, r_0}$ with $r_0 = 0.5$,
  and with a L\'evy flight motion dynamics, obtained by using
  Eq. \ref{eq:lf} with $\gamma = 2.5$ for extracting the step length.  }
\end{figure}

The model discussed above depends on different numerical and functional
parameters, namely the individual density $\rho$, the attractiveness
distribution $\eta(a)$ and the activation probability distribution
$\phi(r)$.  As we have seen, some properties of the model, especially
those related to group and collective level dynamics, do indeed depend
of the density $\rho$ (or the number of individuals $N$), in such a way
that model is able to reproduce empirical data only when fed with a
value of $N$ corresponding to the data set under consideration.  The
model properties relevant to the individual level dynamics however, such
as the contact duration and weight distributions, $P(\Delta t)$ and
$P(w)$, do not change in a reasonable range of density.
In Fig. \ref{fig:robust} one can observe that the functional form of
these distributions is robust with respect to changes of the individual
density, supporting the natural notion that individual level dynamics is
mainly determined by close contacts of pairs of individuals, and rather
independent of eventual multiple contacts, which become rarer for
smaller densities.

We have also explored the dependence of the model on the activation
probability distribution and the walking probability. In particular,
instead of a uniform activation probability distribution, we have
considered a constant distribution 
\begin{equation}
  \phi(r) = \delta_{r, r_0},
  \label{eq:3}
\end{equation}
where $\delta_{r, r'}$ is the Kronecker symbol. 
 As we can see from Fig. \ref{fig:robust} the output of the model is 
robust with respect to changes of this functional parameter. 


Finally, in the definition of the model we have adopted the simplest
motion dynamics for individuals, namely an isotropic random walk in
which the distance $v$ covered by the agents at each step is constant
and arbitrarily fixed to $v=1$. However, it has been noted for long that
a L\'evy flight turns out to provide a better characterization of human
or animal movement and foraging \cite{viswanathan2011physics}. In this
case, the random walk is still isotropic, but now the distance covered
in each step is a random variable, extracted from a probability
distribution
\begin{equation}
\label{eq:lf}
\mathcal{L}(v) \simeq v ^{-\gamma},  
\end{equation}
with a long tailed form.  In Fig. \ref{fig:robust} we show that adopting
a L\'evy flight motion dynamics gives rise to outcomes in very good
agreement with the original definition of the model.  We note, however,
that the step length probability distribution $\mathcal{L}(v)$ has a
natural cutoff given by the size of the box where the agents move,
$v<L$, reducing the degree of heterogeneity that the walk can cover.

\section{Discussion}
\label{sec:discussion}

Understanding the temporal and structural properties of human contact
networks has important consequences for social sciences, cognitive
sciences, and epidemiology.  The interest in this area is not new, but has been fueled
by the recent availability of large amounts of empirical data, obtained
from expressly designed experimental setups.  The universal features
observed in these empirical studies prompt for the design of general
models, capable of accounting for the observed statistical regularities.

In this paper we reported the results obtained from a simple model in
which individuals perform a random walk and start interactions based on
a close proximity rule.  The key ingredient is the social attractiveness
of the individuals, which has the effect of slowing down the random walk
performed by the agents and determines the duration of their
interactions.  By means of numerical simulations, we observed that the
model reproduces the results obtained from the empirical analysis of the
human contact networks provided by the SocioPatterns collaboration
\cite{sociopatterns}.  The match between the model and the empirical
results is independent of the numerical and functional form of the
diverse parameters defining the model.  However, the attractiveness
distribution $\eta(a)$ used in the model definition deserves a more
detailed discussion.  Its functional form is hard to access empirically,
and it is likely to be in its turn the combination of different
elements, such as prestige, status, role, etc.  Moreover, even though in
general attractiveness is a relational variable -- the same individual
exerting different interest on different agents -- we have assumed the
simplest case of a uniform distribution for the attractiveness.  For
this reason it is important to stress some facts that support our
decision, and to investigate the effect of the attractiveness
distribution on the model outcome.

The choice of a uniform $\eta(a)$ is dictated by the maximum entropy principle, 
according to which the best guess for a unknown but bounded
distribution (as the attractiveness distribution has to be, if we want
it to represent a probability) is precisely the uniform distribution.
However, we can also explore the relation between the
attractiveness and some other variables that can be accessed empirically.  
In particular, the attractiveness of one individual
and the strength of the corresponding node of the integrated network are expected to
be (non-trivially) related, since the more attractive an individual is,
the longer the other agents will try to engage him in interactions.   
Fig. \ref{fig:w_distr} shows that the strength distribution $P(s)$ of
the time-integrated network obtained from the empirical data, which
follows approximately an exponential behavior, is well fitted by the model. 
Thus, if we hypothesize that the attractiveness and strength
probability distributions are related as $P(s) ds \sim \eta(a) da$, with
$\eta(a)$ uniform in [0,1], it follows that the strength of an
individual should depend on his attractiveness as
 \begin{equation}
 \label{eq:s_a}
s(a) \sim - \log (1-a).
\end{equation}
We find that this relation is fulfilled by the model (data not shown),
showing that in the model the time spent in interactions by the individuals is
directly related with their degree of attractiveness. 
Therefore the guess of a heterogeneous but uniform $\eta(a)$ leads to a exponential
decay of the $P(s)$ for the model, in accordance with experimental data,
and providing grounds to justify this choice of attractiveness distribution.
Moreover, the simple relation expressed by Eq. \ref{eq:s_a} 
may suggest a way to validate the model, once some reliable measure of attractiveness will be available.

Finally, it is worth highlighting that the form of the attractiveness
distribution $\eta(a)$ is crucial for the model outcome.  
Therefore, it is interesting to explore the effect of different functional forms of 
$\eta(a)$, for example incorporating a higher degree of heterogeneity,
such as in the case of a power-law distribution.  We note, however, that
the form of Eq. (\ref{eq:mr}) imposes to the $a_i$ variable to be a
probability, with the consequent constraint of being bounded, $a_i \in [0,1]$.  
A power law distribution defined over a bounded support presents some issues, 
such as the necessity of imposing a lower bound to prevent divergence close to $0$.  
To avoid this inconvenient, one can redefine the motion rule of Eq. (\ref{eq:mr}) as
\begin{equation}
\label{eq:mr_inverse}
p^{\rm{inv}}_i(t) = \frac{1}{\max_{j \in \mathcal{N}_i(t) } \{ a'_j \} },
\end{equation} 
with the new attractiveness variable $a'_i$ unbounded,  $a'_i \in [1, \infty)$.
If we impose the walking probability of Eqs. (\ref{eq:mr}) and
(\ref{eq:mr_inverse}) to be the same, and we use the relation $\eta(a)
da = \zeta(a') da'$, we find that the new attractiveness distribution
$\zeta(a')$ has the form of a power law, $\zeta(a') = (\gamma -1 )
a'^{-\gamma}$, with exponent $\gamma = 2$.  Therefore, assuming a motion
rule of the form of Eq. (\ref{eq:mr_inverse}), a power law
attractiveness distribution will give rise to same model results, as
confirmed by numerical simulations (data not shown).

On the same line of argument, it would be interesting to relate the
agents' activation probability, $r_i$, with some empirically accessible
properties of the individuals. Unfortunately, finding the activation
probability distribution $\phi(r)$ is a hard task with the information
contained in the available datasets.  In the face-to-face interaction
deployment, indeed, a non-interacting but active individual is
indistinguishable from an inactive individual who is temporary not
involved into the event.  Thus, simply measuring probability to be not
involved in a conversation does not inform on the $\phi(r)$, but instead
considers something more related with the burstiness of the individual
activity.  In any case, however, the model behavior is independent of
the functional form the activation distribution, so that this point is
less crucial.

In summary, we showed that a simple model based on the concept of social
attractiveness is able to account for the main statistical properties of
human contact networks at different scales. This finding prompts for
further empirical research, based on more detailed and extensive
experimental setups, which can shed light on the role of this
attractiveness. Such research would help to further refine and validate
the model considered here, and could potentially provide new insights
for the social and cognitive sciences.

\section{Acknowledgments}

We acknowledge financial support from the Spanish MEC, under project
FIS2010-21781-C02-01, and EC FET-Proactive Project MULTIPLEX (Grant
No. 317532).  We thank the SocioPatterns collaboration for providing
privileged access to datasets ``Hospital'', ``School'', and
``Congress''. Dataset ``Conference'' is publicly available
at~\cite{sociopatterns}.

\bibliographystyle{abbrv}


\bibliography{attractiveness}

\end{document}